\documentclass[prl,aps,twocolumn,superscriptaddress,tightenlines,floatfix]{revtex4}
\usepackage{graphicx}
\usepackage{dcolumn}
\usepackage{amsmath}

\begin{document}
\title{Temperature dependence of the spin and orbital magnetization density in $Sm_{1-x}Gd_{x} Al_{2}$ around the spin-orbital compensation point.}
\author{J.W. Taylor}
\affiliation{Department of Physics, The University of Warwick,
Coventry, CV4 7AL, UK.}
\author{J.A. Duffy}
\affiliation{Department of Physics, The University of Warwick,
Coventry, CV4 7AL, UK.}
\author{A.M. Bebb}
\affiliation{Department
of Physics, The University of Warwick, Coventry, CV4 7AL, UK.}
\author{M.R. Lees}
\affiliation{Department of Physics, The University of Warwick,
Coventry, CV4 7AL, UK.}
\author{L. Bouchenoire}
\affiliation{Department of Physics, The University of Warwick,
Coventry, CV4 7AL, UK.}
\affiliation{XMaS, The UK CRG, European
Synchrotron Radiation Facility, BP 220, F-38043, Grenoble Cedex,
France.}
\author{S.D. Brown}
\affiliation{XMaS, The UK CRG, European Synchrotron Radiation
Facility, BP 220, F-38043, Grenoble Cedex, France.}
\author{M.J. Cooper}
\affiliation{Department of Physics, The University of Warwick,
Coventry, CV4 7AL, UK.}
\date{\today}
\begin{abstract}
Non-resonant ferromagnetic x-ray diffraction has been used to
separate the spin and orbital contribution to the magnetization
density of the proposed zero-moment ferromagnet
$Sm_{0.982}Gd_{0.018} Al_{2}$. The alignment of the spin and
orbital moments relative to the net magnetization shows a sign
reversal at 84K, the compensation temperature. Below this
temperature the orbital moment is larger than the spin moment, and
vice versa above it.  This result implies that the compensation
mechanism is driven by the different temperature dependencies of
the $4f$ spin and orbital moments. Specific heat data indicate
that the system remains ferromagnetically ordered throughout.
\end{abstract}
 \maketitle

Recently, it was proposed that the Laves phase compound
$Sm_{1-x}Gd_{x} Al_{2}$ exhibits a spin-orbital compensation point
at $\approx85K$ when x = 0.0185 \cite{adachi:99}. Magnetometery
showed that the net moment dipped to zero at this temperature, but
was finite either side in the magnetically ordered phase (the
Curie temperature is 128K). At the compensation temperature,
magnetic Compton scattering shows a net {\it spin} moment,
indicating that the system consists of a ferromagnetically ordered
spin sublattice\cite{adachi:01}. For the net moment to be zero,
this spin moment must be exactly compensated by the orbital
moment. Although it is thought that this may be driven by the
different temperature dependencies of the Sm $4f$ spin and orbital
moments, this had not yet been investigated. An understanding of
the compensation mechanism may be gained by studying the
temperature dependence of the spin and orbital moments near the
compensation point, which requires direct measurement of the spin
and orbital magnetization. Such an x-ray diffraction study is
reported here for the first time. Our result conclusively proves
that the compensation point is driven by the different temperature
dependence of the spin and the orbital moments. Our specific heat
data indicate that the system remains magnetically ordered.

The magnetism of Sm and its compounds has been the focus of many
investigations as a result of the importance of the conduction
electron polarization and the complex crystalline electric field
(CEF) in the material\cite{stewart:72}. The spin and orbital
contribution to the $Sm^{3+}$ $4f$ moment are of similar size and
aligned antiparallel, and the polarized conduction electron spin
moment is thought to align parallel with the $4f$ spin
moment\cite{adachi:97r}. The three components to the site
magnetization almost cancel, leaving a small net local moment.
Interestingly, the temperature dependencies of the spin and
orbital components are not identical due to a complex thermal
admixture of nearly degenerate \textbf{J} multiplets in which the
$Sm^{3+}$ ion exists in (the ground state $5/2$ muliplet is 1500K
from first excited state $7/2$). The admixture arises from the CEF
effect on the degeneracy of the \textbf{J} states and has long
been an explanation of the magnetism in Sm compounds
\cite{malik:71}.

A solid solution of $Gd^{3+}$, introduces a large (7.6 $\mu_{B}$)
spin moment onto the Sm site in $Sm_{1-x}Gd_{x} Al_{2}$ and the
small induced lattice distortion alters the CEF deformation
potential. It also critically affects the RKKY exchange
interaction due to the increase in conduction electron
polarization. These factors have a considerable influence on the
temperature dependence of the Sm site moment, as the thermal
admixture of \textbf{J} states is re-normalized. The result is to
change the temperature dependencies of the $4f$ spin and orbital
moments. In the undoped compound the moments are $M_{l} \approx
4.3\mu_{B}$ and $M_{s}\approx-3.8\mu_{B}$ \cite{adachi:99a}
respectively. The change in the CEF allows the Sm orbital and
Gd/Sm spin contributions to cancel each other completely at a
distinct temperature below $T_{C}$:
at this point the material has no net moment, and is referred to
as compensated. This effect in itself is not unusual in some
ferrimagnetic systems, where \textit{two} sublattice
magnetizations become equal and opposite at a particular
temperature. However in this case the magnetism exists only on the
rare earth site (a solid solution of Sm / Gd ions). A naive
picture of the temperature dependence has three order parameters,
$4f$ orbital magnetism, $4f$ spin magnetism, and conduction
electron spin polarization with the latter probably having the
same temperature dependence as that of the $4f$ spin. If the order
parameters are of opposite sign, with non-identical temperature
dependence the system can become compensated. Previous work has
concentrated on the bulk magnetization and the type of magnetic
ordering at the compensation point. However the mechanism of
compensation in $Sm_{1-x}Gd_{x} Al_{2}$ has not been investigated.

In this letter we report the use of non-resonant x-ray diffraction
to investigate the magnetization density of $Sm_{0.98}Gd_{0.012}
Al_{2}$ as a function of temperature through the spin-orbital
compensation point by monitoring a Bragg reflection. The technique
has the advantage of allowing the separation of the spin and
orbital form factors by changing the experimental geometry. At the
wave-vector sampled, the conduction electron moment makes no
contribution to the form factor, therefore the technique gives
direct access to the temperature dependence of the spin and
orbital components of the $4f$ moment.

This is the first direct observation of the temperature dependence
of the $4f$ spin and orbital form factors. From our investigation
it is clear that the conduction electron polarization is a
critical factor in the compensation process as, at $T>T_{comp}$,
the $4f$ moments remain almost compensated. From our own
investigation of the bulk properties of the material, in
particular the magnetic contribution to the specific heat,
presented later, it is clear that at the compensation point the
material exhibits no change in entropy, and therefore does not
exhibit any sign of a magnetic transition. This is due, in part,
to the high anisotropy in the system.

The technique of non-resonant x-ray diffraction has recently been
developed as a convenient method of studying the spin density in
ferromagnetic materials using elliptically polarized synchrotron
radiation.
\begin{figure}
\begin{center}
\includegraphics[width=8cm]{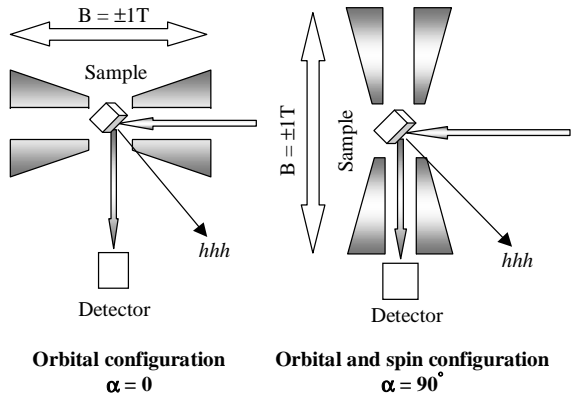}
\end{center}
\caption{Schematic layout of a generic non-resonant ferromagnetic
diffraction experiment. Left hand figure shows experimental
configuration for measurement of the orbital component of the form
factor. Right hand figure shows configuration for measurement of
the total (L+S) form factor.}\label{1}
\end{figure}
It is particularly useful for materials where the neutron
technique is not viable due to the high neutron absorption cross
section. The technique also gives a convenient method  of
separating the spin and the orbital contribution to the total
magnetic form factor by changing the experimental geometry (see
fig~\ref{1}). Essentially the technique makes use of the
suppression of Thompson charge scattering at a scattering angle of
$90^{\circ}$ for radiation linearly polarized in the scattering
plane. When elliptically polarized photons are incident, the
charge and magnetic Bragg intensities interfere. This leads to
modulation of the signal with reversal of the sign of magnetic
component, this can be achieved by either flipping the sample
magnetization vector (in the scattering plane), or by flipping the
helicity of the incident beam polarization. When the pure charge
scattering is a minimum, the signal modulation, resulting from the
magnetic scattering cross section, tends to a maximum, which
facilitates the measurement of a flipping ratio.

The fractional change in intensity upon reversal of the sample
magnetization or the photon helicity is related to the orbital
($F_{L}$), spin ($F_{S}$) and charge($F_{C}$) form factors as:

\begin{equation}
R(\alpha) =
\frac{I_{\uparrow}-I_{\downarrow}}{I_{\uparrow}+I_{\downarrow}}
=gf_{p} \frac{2F_{S}(\textbf{k})sin\alpha +
F_{L}(\textbf{k})(sin\alpha+cos\alpha)} {F_{C}(\textbf{k})}
\label{eq1}
\end{equation}
where $g=\hbar\omega / m_{e}c^{2}$, $f_{p}=P_{c} / (1-P_{l})$,
$P_{c}$ and $P_{l}$ are parameters for the circular and linear
component of the beam, and thus describe the degree of ellipticity
of the incident photons.

The orbital and spin contributions to the form factor are defined
in terms of $\alpha$ the angle between the B field and the
incident beam (see fig 1):

\begin{equation}
R(\alpha = 0) = -2gf_{p} \frac{F_{L}}{F_{C}}
\label{eq2}
\end{equation}

\begin{equation}
R(\alpha = 90^{\circ}) = -2gf_{p} \frac{F_{S}+F_{L}}{F_{C}}
\label{eq3}
\end{equation}
In the past most experiments have made use of a polychromatic
incident beam of x-rays, in order to collect data on a number of
Bragg peaks simultaneously, using energy dispersive Ge detectors.
However, the white beam method suffers from multiple diffraction
which corrupts the signal and is difficult to model. In this
investigation a monochromatic beam was used to avoid these
uncertainties. However the principle remains identical to that
described previously for white beam
experiments\cite{collins:93},\cite{laundy:98}.

A single crystal sample of $Sm_{1-x}Gd_{x} Al_{2}$ with $x=0.018$
was produced by the Bridgemann method, with the polycrystalline
boule sealed in a Ta can to maintain stoichiometry. The structure
of the resulting crystal was verified as the $C_{15}$ Laves phase
using Laue photography.

The non resonant magnetic diffraction experiment was performed on
the XMaS beamline\cite{xmas:01} at ESRF. Elliptically polarized
radiation was extracted by viewing the bending magnet source at a
angle of $\approx$ 0.3mrad from the plane of the synchrotron: the
optimum position in terms of signal to noise.
The sample was mounted in a Be shrouded closed cycle He cryostat.
An incident energy of 5.736keV was selected using the double
bounce Si monochromator, such that the 333 reflection was in the
Bragg condition with a scattering angle of 90$^{\circ}$ in the
plane of the synchrotron (see fig 1). The calculated polarization
of the beam at the incident energy used was $P_{l}= 0.99470$ and
$P_{c} = -0.02937$ \cite{laudny:prog}, yielding a polarization
factor, $f_{p}$, of -5.5. The diffracted beam was detected using a
fast NaI scintillator with an average count rate of $\approx$
85000cps at the diffraction peak. The magnetic field was applied
using a 1T electromagnet, which was flipped at intervals of 20s in
order to average over the beam position fluctuations inherent with
the bending magnet source. In this configuration a single flipping
ratio was acquired over an integration time of 2hours. The
flipping ratios of the 333 reflection were measured as a function
of temperature, in the total (eq 3) moment configuration and the
orbital only (eq 2) configuration, in both heating and cooling
cycles to ensure reproducibility of the data. At $sin \theta /
\lambda$ = 0.32$\AA^{-1}$ on the form factor curve only the $4f$
moments contribute to the magnetic signal.

A comprehensive investigation of the magnetic properties of the
sample was performed at Warwick University using SQUID and VSM
magnetometery, specific heat measurements and AC susceptibility,
in order to check sample quality and to investigate the complex
magnetic properties of the sample comprehensively.

\begin{figure}
\begin{center}
\includegraphics[width=8.5cm]{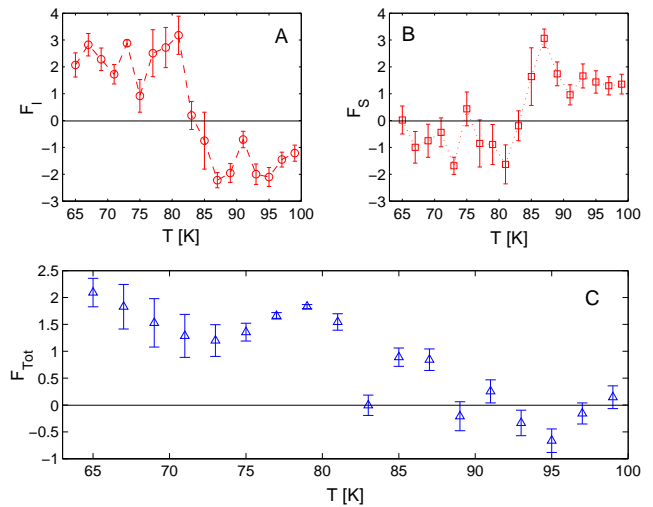}
\end{center}
\caption{Temperature dependence of the flipping ratio of the 333
reflection for $Sm_{1-x}Gd_{x} Al_{2}$. \textbf{A}: Orbital only
form factor temperature dependence (Circle). \textbf{B}: Derived
spin form factor temperature dependence(Square). \textbf{C}:
Temperature dependence of the total form factor
(Triangle).}\label{2}
\end{figure}

For reflections of the type $hhh$ where h is odd, only the $4f$
site contributes to the phase factor. The temperature dependence
of the orbital, spin and total form factor curve at $sin\theta /
\lambda$ = 0.32$\AA^{-1}$ is shown in fig~\ref{2}. It is clear
that below the compensation temperature the orbital contribution
is positive (fig~\ref{2} A), and thus the derived spin
contribution is negative (fig~\ref{2} B), with a smaller magnitude
as expected, since at this wave vector the conduction electron
polarization is not measured. The spin-only form factor result is
in good agreement with that measured previously for the un-doped
sample\cite{adachi:01a}. Above the compensation temperature the
spin and orbital contributions are reversed, with approximately
equal magnitudes. The total form factor (fig~\ref{2} C) is
positive below the compensation temperature, as expected, since
the total magnetization will follow the large orbital
contribution. Interestingly above $T_{comp}$ the total form factor
is negligible. It is clear that both the orbital and the spin
component to the form factor have a complex temperature
dependence, furthermore both components flip sign at $T_{comp}$,
with the total form factor tending to zero at $T_{comp}$. This
result implies that the systems exhibits no ferro-magnetic
character at $T_{comp}$. This does not mean that the system
becomes paramagnetic however, or that the orbital, or spin
magnetizations disappear at the compensation point, for the
following reasons.

The orbital and spin components to the magnetization have
different temperature dependence and an anti-parallel arrangement.
The net moment in the system will always align with the field,
(when the field is large i.e. 1T see next section). At low
temperature this results in a positive contribution to the
magnetization density arising from the orbital moment and a
negative contribution arising from the spin moment, with a net
positive magnetization density where L$>$S. When the system
becomes compensated the orbital and spin components are in effect
antiferro-magnetically aligned. However the orbital and the spin
components should still be finite. Our data show a definite sign
reversal of the spin and the orbital components. The change of
sign results from the spin component becoming dominant above
$T_{comp}$, hence the net moment is re-aligned with respect to the
field.

The fact that our orbital data tend to zero smoothly at $T_{comp}$
rather than exhibiting a sharp step-like transition, is a
statistical artefact produced from a combination of unwanted beam
movements, from the synchrotron bending magnet source, and
temperature fluctuations in the cryostat due to reversing the
applied field. If one assumes that the temperature is only stable
to within $\pm 0.5K$ one may easily reach a point whereby the
sample is driven from one side of $T_{comp}$ to the other, by the
eddy current heating effect, throughout the period of the
measurement, thereby measuring zero.

Above $T_{comp}$ the measured form factor is negligible. This
means that the $4f$ components to the magnetization are of a
similar size, which in turn implies that the conduction electron
spin component (not measured by the diffraction experiment) is of
critical importance. Our diffraction data provide clear evidence
that the spin and the orbital contributions to the $4f$
magnetization density cancel at the compensation point, and that
the compensation point occurs as a result of the different
temperature dependence of the spin and orbital form factor.

\begin{figure}
\begin{center}
\includegraphics[width=8.9cm]{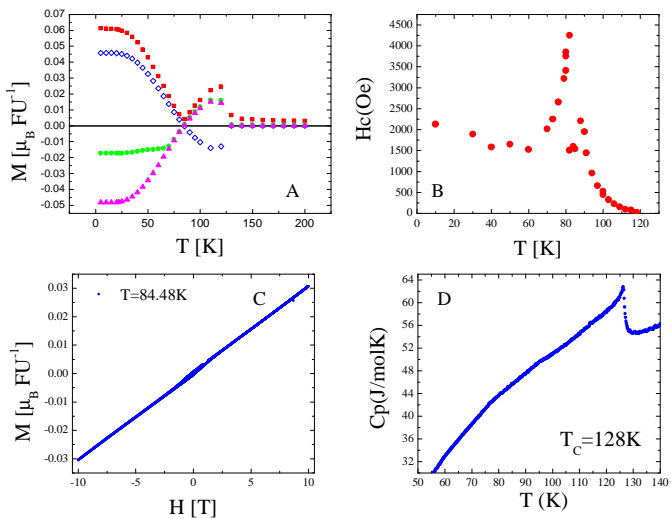}
\end{center}
\caption{Low temperature properties of $Sm_{1-x}Gd_{x} Al_{2}$.
\textbf{A}: Magnetization as a function of temperature. Triangles
Field cooled in $10^{-2}$T. Circles field cooled in $0.1$T.
Diamonds zero field cooled. Squares field cooled in $1$T.
\textbf{B}: Temperature dependence of the coercive field.
\textbf{C}: Magnetization as a function of field (up to 10 T) at
$T_{comp}$. \textbf{D}: Specific heat capacity.}\label{3}
\end{figure}

This diffraction result will now be discussed in the context of
the bulk properties of the system. Firstly the magnetization data
observed for the sample as a function of temperature
(fig~\ref{3}:A). The bulk magnetization data clearly show that, at
the compensation point, the net moment in the sample is zero.
However the magnetic behavior below $T_{comp}$ drastically alters,
depending on whether the sample is field cooled or zero field
cooled. On cooling in a small $10^{-2}$T field the magnetization
shows a large diamagnetic effect below $T_{comp}$ (fig~\ref{3}:A,
triangles). The size of the diamagnetic effect can be altered by
changing the magnitude of the applied field (fig~\ref{3}:A, $0.1$T
(circles) and $1$T (squares)). If the system in first field cooled
and then warmed through $T_{comp}$ in the remnant field the system
again shows some diamagnetic effect.

This is an effect of the large anisotropy within the system. The
Sm $4f$ moment can be thought of as a single site ferri-magnet,
with the spin and orbital contributions having different
temperature dependencies if the size of each contribution is
reversed either side of $T_{comp}$ the system must overcome the
magnetocrystalline anisotropy energy (MAE) to realign the net
moment with the applied (or remnant) field as is observed in the
diffraction experiment. It follows that if $\mu.B<MAE$ the system
exhibits a large diamagnetic effect. The strong MAE effect in this
system is demonstrated by the temperature dependence of the
coercive field (fig~\ref{3}:B) which clearly shows maxima above
and below $T_{comp}$. When considered with the diffraction data
the behavior of the magnetic system is clarified. The diffraction
data were taken in an applied field of 1T, which is large enough
to overcome the large MAE and hence realign the net moment in the
system. The closed hysteresis loop at $T_{comp}$ (fig~\ref{3}:C)
is verification that the system is compensated and indicating a
strongly correlated quasi anti-ferromagnetic behavior

The specific heat capacity (fig~\ref{3}:D) shows a large
$\lambda$-type anomaly associated with the Curie temperature at
$T_{c}$=128K, however there is no effect in the specific heat data
at a temperature corresponding to $T_{comp}$ which indicates that
there is no transition at that temperature.

It is obvious that the conduction electron moment is an important
factor in the magnetization in this material. The alloying of
$2\%$ Gd will certainly affect the RKKY polarization in the system
due to the lattice distortion and also the large size of the Gd
moment. It is also clear that the $4f$ contributions above
$T_{comp}$ appear, from our data to be equal. Thus it is
reasonable to assume that the net magnetization observed above
$T_{comp}$ (fig~\ref{3}:A) results almost exclusively from the
conduction electron moment in the system, the temperature
dependence of which is unknown (although it is reasonable to
assume it is similar to the $4f$ spin moment) Such a measurement
is planned using the magnetic Compton scattering technique, which
directly samples the polarization of all spin polarized electrons.

In conclusion, our data show that the total $4f$ magnetization
density is zero at the compensation temperature. We have
demonstrated that the compensation mechanism is driven by the
temperature dependence of the spin and orbital moments in the
system. We have shown that the unusual temperature dependence of
the bulk magnetization is driven by the reversal of the dominant
$4f$ component at the compensation temperature, i.e.
$T<T_{comp}$:L$>$S and $T>T_{comp}$:S$>$L. The fact the bulk
measurement of specific heat shows no anomaly at the $T_{comp}$
implies that magnetic system remains ordered, as one may expect
due to the high magnetocrystalline anisotropy energy.

This work was performed on the EPSRC funded beamline XMaS at ESRF.
The authors thank Dr D. Fort of the University of Birmingham for
growing the single crystal sample used in this investigation, and
Dr D. Mannix and other beamline staff for their assistance. The
authors thank EPSRC(UK) for funding.


\begin{thebibliography}{99}

\bibitem{adachi:99}
H. Adachi and H. Ino, Nature, \textbf{401}, 148 (1999).
\bibitem{adachi:01}
H. Adachi, H. Kawata, H. Hashimoto, Y. Sato, I. Matsumoto and Y.
Tanaka, Phys. Rev. Lett., \textbf{87} 127202 (2001)
\bibitem{stewart:72}
A.M. Stewart Phys. Rev. B \textbf{6} 1985 (1972)
\bibitem{adachi:97r}
H. Adachi, H. Ino, A. Koizumi, N. Sakai, Y. Tanaka and H. Kawata,
Phys. Rev. B \textbf{56}, R5744 (1997)

\bibitem{malik:71}
S.K. malik and R. Vijayaraghavan, Phys. Lett. \textbf{34} 67
(1971).

\bibitem{adachi:99a}
H. Adcahi, H.Ino and H. Miwa, Phys. Rev. B \textbf{59} 11445
(1999).

\bibitem{collins:93}
S.P. Collins, D. Laundy and G.Y. Guo, J. Phys. Condens.:Matter
\textbf{5} L637 (1993).
\bibitem{laundy:98}
D. Laundy, S. Brown, M.J. Cooper, D. Bowyer, P. Thopmpson, W.G.
Stirling and J.B. Forsyth, J. Synch. Rad. \textbf{5} 1235 (1998).
\bibitem{xmas:01}
S.D. Brown, L. Bouchenoire, D. Bowyer, J. Kervin, D. Laundy, M.J.
Longfield, D. Mannix, D.F. Paul, A. Stunault, P. Thompson, M.J
Cooper, C.A. Lucas and W.G. Stirling,  J. Synch. Rad., \textbf{8}
18 (2001).

\bibitem{laudny:prog}
Calculated using the program written by D. laundy. SRS UK.
\bibitem{adachi:01a}
H.Adachi H. Kawata and M. Ito, Phys. Rev. B. \textbf{63} 054406
(2001).
\end{thebibliography}
\end{document}